\documentclass[fleqn,usenatbib]{mnras}
\usepackage{graphicx}
\usepackage{graphics}
\usepackage{amsmath, amssymb}
\usepackage{multirow}
\usepackage{color}
\usepackage{xcolor}
\usepackage{bm}		% Bold maths symbols, including upright Greek
\usepackage[normalem]{ulem}

\usepackage{float}
\def\mathnew{\mathsurround=0pt}   
\def\simov#1#2{\lower .5pt\vbox{\baselineskip0pt  
    \lineskip-.5pt\ialign{$\mathnew#1\hfil##\hfil$\crcr#2\crcr\sim\crcr}}}

\def\'#1{\ifx#1i{\accent"13\i}\else{\accent"13#1}\fi}

\def\aap{{A\&A}}

\def\aj{{AJ}}
\def\apj{{ApJ}}
\def\apjl{{ApJL}}
\def\nat{{Nature}}
\def\mnras{{MNRAS}}
\def\apjs{{ApJS}}

\def\rmxaa{{Rev. Mex. Astron. Astrofis.}}

\title[Ridges in velocities \& wiggles in RCs]{From ridges in the velocity distribution to wiggles in the rotation curve}

\author[Martinez-Medina et al. 2019]{
Luis Martinez-Medina$^{1,2}$\thanks{Contact e-mail:\href{mailto:lamartinez@astro.unam.mx}{lamartinez@astro.unam.mx}}, 
Barbara Pichardo$^1$, 
Antonio Peimbert$^1$, 
Octavio Valenzuela$^1$\\
$^1$ Instituto de Astronom\'ia, Universidad Nacional Aut\'onoma de M\'exico, A.P. 70--264, 04510, M\'exico, CDMX, M\'exico.\\
$^2$ Department of Physics, University of Surrey, Guildford, GU2 7XH, UK.\\
}

\date{Accepted 2019 March 20; in original form 2019 January 17}

\pubyear{2019}

\begin{document}
\label{firstpage}
\pagerange{\pageref{firstpage}--\pageref{lastpage}}
\maketitle

\begin{abstract}
Recently, the Gaia data release 2 (DR2) showed us the richness in the kinematics of the Milky Way disk. Of particular interest is the presence of ridges covering the stellar velocity distribution, $V_{\phi}-R$; as shown by others, it is likely that these ridges are the signature of phase mixing, transient spirals, or the bar. Here, with a Galactic model containing both: bar and spirals, we found the same pattern of ridges extending from the inner to the outer disk. Interestingly, ridges in the $V_{\phi}-R$ plane correlate extremely well with wiggles in the computed rotation curve (RC). Hence, although the DR2 reveals (for the first time) such substructures in a wide spatial coverage, we notice that we have always seen such pattern of ridges, but projected into the form of wiggles in the RC. The separation and amplitude of the wiggles strongly depend on the extension and layout of ridges in the $V_{\phi}-R$ plane. This means that within the RC are encoded the kinematic state of the disk as well as information about the bar and spiral arms. The amplitude of the wiggles suggests that similar features currently observable in external galaxies RCs have similar origins, triggered by spirals and bars.

\end{abstract}                
 
 
\section{Introduction} \label{sec:intro}

The second Gaia data release (DR2) has come with an unprecedented amount of information on the kinematics of the Galactic disk. The various phase space planes are full of substructures: arches, shells and spirals in the U-V-W planes \citep{2018arXiv180410196A}, and ridges in the $V_{\phi} - R$ plane \citep{2018arXiv180410196A,2018MNRAS.479L.108K,2018arXiv180509790R}. As already pointed out, the rolling up spiral in phase space, reveals that the MW disk is in a clear phase mixing stage. Qualitatively, some of the kinematic structure can be reproduced in models with a central bar \citep{2018arXiv180410196A}, transient spiral arms \citep{2018MNRAS.481.3794H}, or a spiral pattern excited by the interaction of the disk with a dwarf galaxy \citep{2018arXiv180800451L}.

This detailed richness of the stellar disk kinematics reveals a more complex stage, and a more complex Galactic evolution, than previously thought. However, the unprecedented precision and completeness of the Gaia DR2 also guides the way and points the direction to better describe and understand our Galaxy. Recent analysis of the DR2 phase space substructures focus mainly on the stellar U-V plane, finding that the data can be explained by transient spiral arms perturbing the MW disk \citep{2018MNRAS.480.3132Q,2019MNRAS.484.3154S}.

However, the kinematics is not only affected by the spirals; it is well known that the bar has a crucial role in shaping the kinematics of the stellar disk as seen in the U-V-W planes \citep{2014A&A...563A..60A,2017ApJ...840L...2P,2018arXiv180401920H}. Probably the takeaway from all the previously mentioned studies is that: it is the combined effect of bar and spirals what sets up the substructures in the kinematics of the MW disk.

Although the U-V-W planes have been extensively explored, the attention on the $V_{\phi}-R$ plane is just recent \citep{2018MNRAS.481.3794H,2019arXiv190107568F}. A deeper exploration of the kinematics projected in this plane surely can provide more insight on the secular evolution of the disk and its perturbers. Furthermore, it is  the $V_{\phi}-R$ plane from where the average rotation of the disk is extracted (i.e. the rotation curve, RC), such that its clumpy nature can propagate to other observables in ways not noticed before.

Then, whereas high-resolution kinematics, similar to the one provided by Gaia, is not observable for other galaxies, it can be inferred from the RC, which is possible to measure with high precision in many galaxies. Of particular interest are the bumps and wiggles in RCs, which have been measured in many disk galaxies with increasing resolution over the past decades \citep{Kalnajs1983,1986AJ.....91.1301K,1996MNRAS.281...27P,2000AJ....120.2884P,2019MNRAS.482.5106L}; and, of those, the most relevant and accurate is the RC of the Milky Way \citep{2014ApJ...783..130R,2019ApJ...870L..10M,Kawata_RC}.

In this work we explore how well the kinematic substructures, in the form of ridges in the $V_{\phi}-R$ plane, are imprinted in the RC. To do that, we analyze the kinematics of a stellar disk evolving in a MW model that contains a bar and spiral arms as non-axisymmetric perturbers.

In section \ref{model} we describe the Galactic model that contains bar + spiral arms. Then, in section \ref{sec:ridges} we analyze in detail the substructures in the $V_{\phi}-R$ plane, comparing them with the Gaia DR2, and describing how they are projected on the RC. Finally we summarize our findings in section \ref{summary}.

\section{Galactic model and simulations}
\label{model}

Our description of the MW disk is based on our own implementation of the chemo-dynamics of the stellar disk. It couples a mass model of the Galaxy with a chemical evolution model of the disk.

First, the Galactic mass model is built by adding the gravitational potential of individual galactic components in order to account for the total Galactic potential and mass. It contains a spherical bulge, a Miyamoto-Nagai disk, and a spherical dark matter halo \citep{AS91}. 

Then, to account for a realistic spiral density and mass distribution, we use the grand design spirals by \citet{PMME03}. It is a careful overlapping of individual inhomogeneous oblate spheroids, in enough number to achieve a smooth but concentrated mass distribution along a logarithmic bisymmetric spiral. For the bar we use a triaxial inhomogenious ellipsoid as described in \citet{Pichardo2004}, which incorporates the smooth density fall found in the COBE/DIRBE data from the galactic center \citep{1998ApJ...492..495F}. Bar and spirals rotate with pattern speeds $\Omega_B = 50kms^{-1}kpc^{-1}$ and $\Omega_S = 25kms^{-1}kpc^{-1}$, respectively, and are introduced adiabatically to the simulation following the smooth transition described by \citet{2000AJ....119..800D}. Discussion about the orbital self-consistency of the model is presented by \citet{PMME03,Pichardo2004}.

An important property of these components is that they represent actual physical densities, in contrast to simpler implementations of bar and spirals commonly used in other works. Such property is an important one because it is a realistic approach to represent the perturbers of the disk, in similarity with the substructures formed in N-body models.

Within this mass model we evolve a disk of test-particles. The disk starts with very few particles and, as the simulation runs, more particles are introduced to it using the star formation rate (SFR) and [Fe/H] evolution obtained from the MW's chemical evolution model by \citet{2011RMxAA..47..139C}. This guarantees consistency between the number and distribution of particles and the chemical distribution. At the end we have the kinematics, age, and metallicity of each particle. Further details on the mass model, chemical evolution model, implementation, and value of the parameters can be found in \citet{2017MNRAS.468.3615M}.
%=======================================================================
\section{Ridges in the $V_{\phi}-R$ plane \& wiggles in the RC}
\label{sec:ridges}

The Gaia DR2 has shown the detailed richness and substructure of the kinematics around the solar neighbourhood. The U-V-W planes have been used to extract information about the nature of the spiral arms \citep{2018MNRAS.480.3132Q,2019MNRAS.484.3154S}, and the $V_Z-Z$ plane gives evidence of phase mixing in the disk \citep{2018arXiv180410196A,2018arXiv181109205K}. Here we focus instead on the diagonal ridges in the $V_{\phi}-R$ plane: a pattern of diagonal over-densities and gaps populating the distribution of rotation velocities of the stellar disk \citep[see figure 2 in][]{2018arXiv180410196A}.

\begin{figure*}
\begin{center}
\includegraphics[width=17cm]{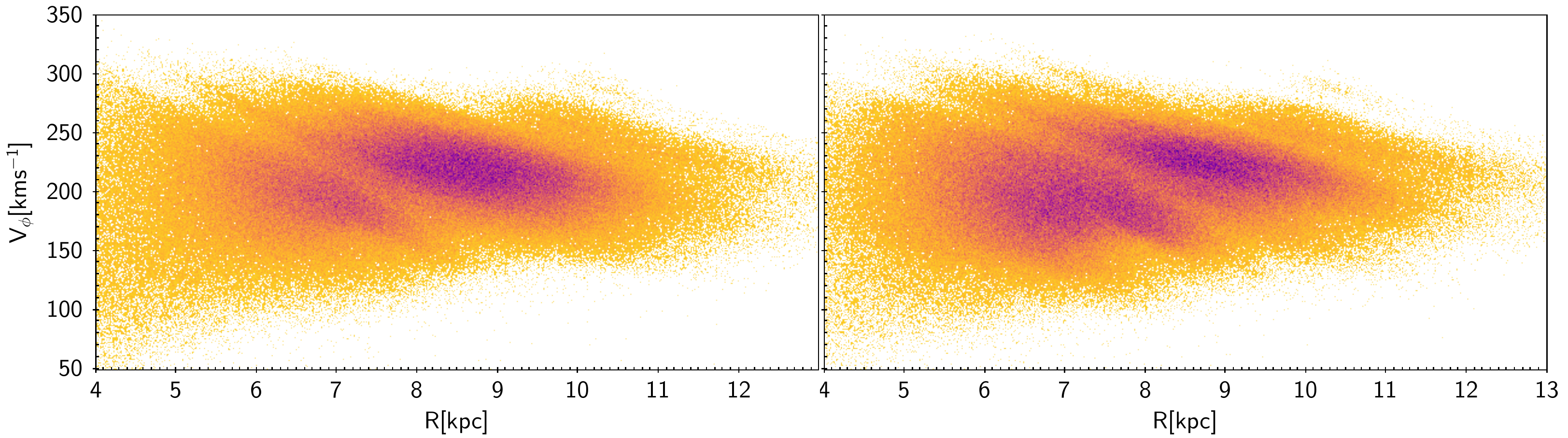}
\end{center}
\caption{Rotation velocity distribution for the stellar disk in our MW simulation. The stars are selected with a gaussian probability around two different points on the solar circle: around the solar neighbourhood (left), and around a point in the perpendicular direction (right). Both $V_{\phi}-R$ planes are populated by wide and thin diagonal ridges, similar to what the Gaia DR2 reveals \citep{2018arXiv180410196A}.}
\label{ridges1}
\end{figure*}

First we compare our Galactic model to the velocity distribution in the DR2. To make a simple but proper comparison with observations we assign to each star a probability of being picked from the simulation. We adopt a gaussian probability centered at the Sun's position and a second one centered 90$^0$ away from the Sun. Figure \ref{ridges1} shows the $V_{\phi}-R$ plane for stars selected around the Sun (left) and for stars selected around another point 90$^0$ away on the solar circle (right). Notice that, in addition to the wide denser ridges, there are many thin ridges, similar to the ones revealed for the first time by Gaia \citep{2018arXiv180410196A}.

We note that the overdensity in the solar neighborhood, observed in the DR2, is much higher than the one produced by the gaussian distribution we use. However, while the Milky Way has $\sim 10^{11}$ stars, our simulation only has $\sim 10^{8}$ stars; and, if we try to represent a more centrally concentrated distribution, what we will end up with is a very poor sampling of the kinematics outside the 8 -- 9 kpc range. Given the similarity of Figure \ref{ridges1} with the substructures in the DR2 we can confirm that the duo bar+spirals plays a leading role in perturbing the stellar disk.

\begin{figure*}
\begin{center}
\includegraphics[width=14cm]{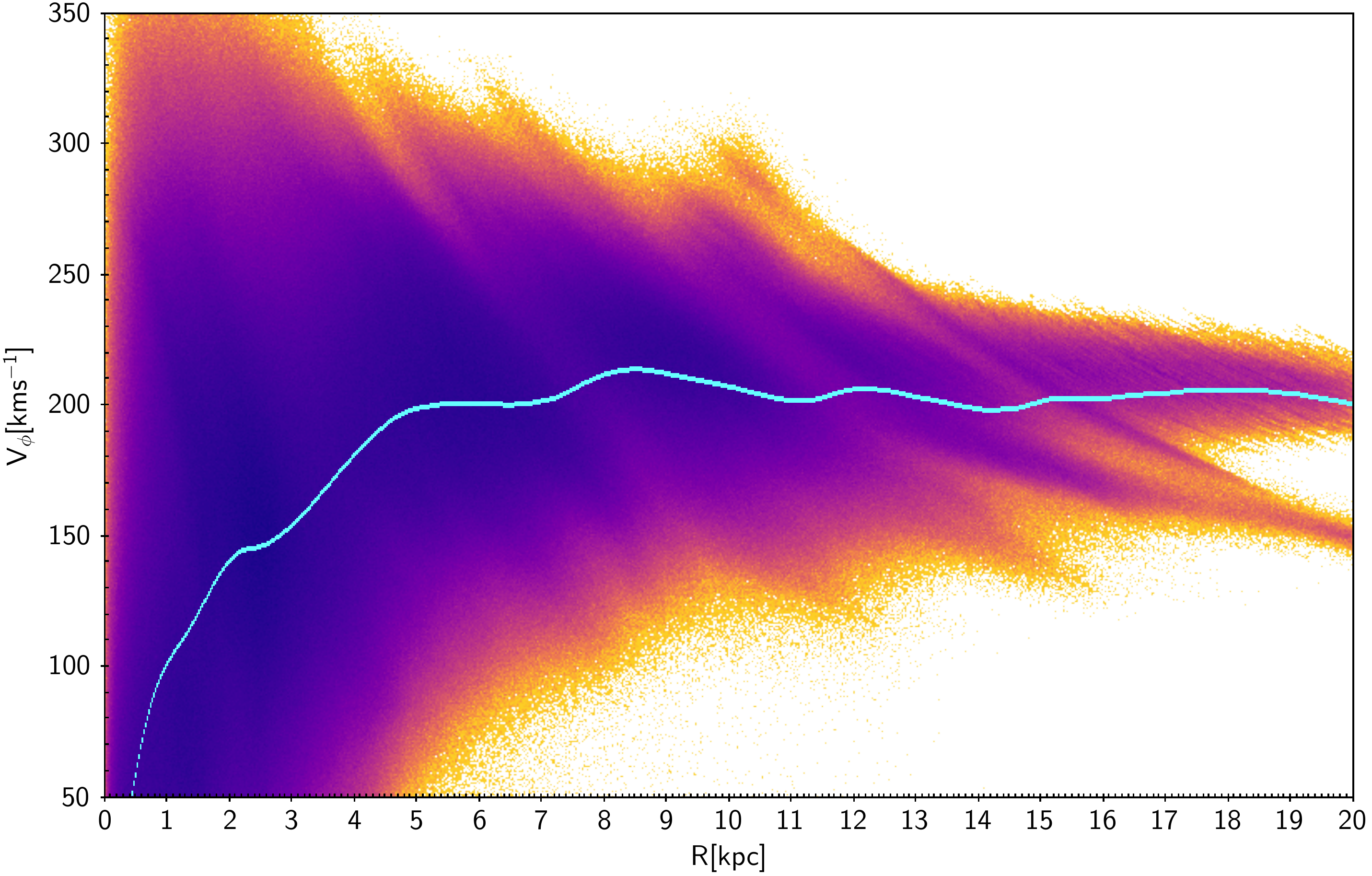}
\end{center}
\caption{Rotation velocity distribution for stars in our MW simulation. The color coding indicates number density (in logarithmic scale) in the $V_{\phi}-R$ plane. Ridges are found across the entire disk. Notice that these diagonal ridges project themselves in the form of wiggles on the RC (blue line).}
\label{ridges}
\end{figure*}

We can now extend our analysis to see how the $V_{\phi}-R$ plane looks like beyond the solar circle. Figure \ref{ridges} shows the stellar velocity distribution for the entire disk. Notice that the pattern of ridges is present along all the extension of the disk, from inside the bar region to the outer disk, beyond the extension of the arms. Given the large spatial coverage of such substructures, the focus is on the implications of their presence in some global properties of the disk, specifically in the RC.

The diagonal nature of such over-densities means that they represent velocity gradients, each ridge begins with high values of $V_{\phi}$, which decrease as the radius increases. As a consequence, if we plot the mean or the median of $V_{\phi}$ in radial bins across the entire disk, the curve will be pulled up near the beginning of each ridge and then pulled down near the end (Figure \ref{ridges}). This produces a sequence of local minima and maxima, giving rise to wiggles, and might be the mechanism responsible for the observed wiggles, which seem to be ubiquitous in RCs \citep{Kalnajs1983,1986AJ.....91.1301K,1996MNRAS.281...27P,2000AJ....120.2884P}.

\begin{figure}
\begin{center}
\includegraphics[width=8cm]{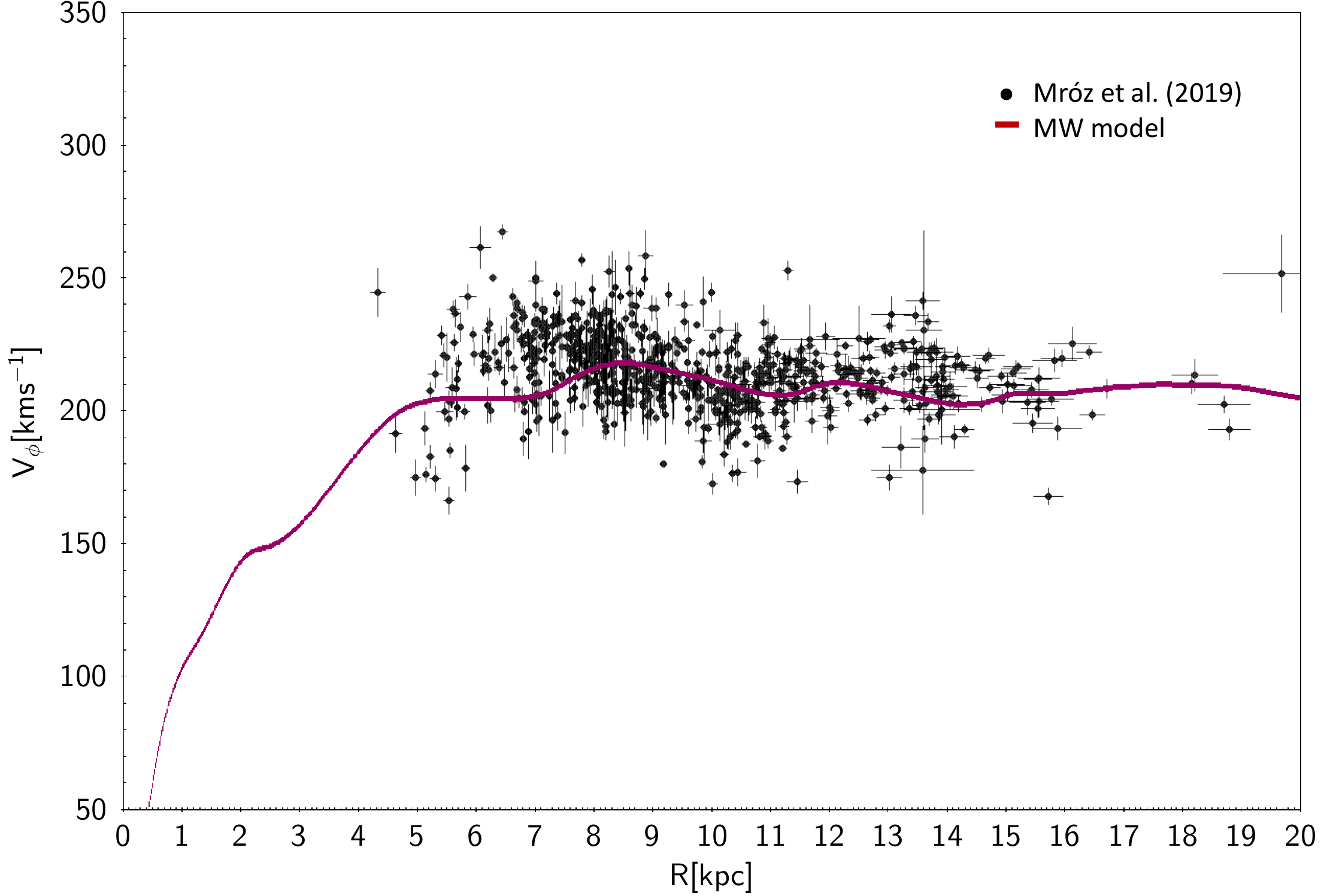}
\end{center}
\caption{MW's RC for cepheids (black dots) constructed from the data by \citet{2019ApJ...870L..10M}. Overplotted is the RC from our MW model, the same displayed in Figure \ref{ridges}.}
\label{MWRC}
\end{figure}

Notice that such correlation between the diagonal ridges in the $V_{\phi}-R$ plane and the wiggles in the RC means that, although most of these features were just revealed by the Gaia DR2, we have always seen them projected in the form of wiggles on the RC. Indeed, Figure \ref{MWRC} shows a comparison between the simulated RC and the MW's RC for cepheids, constructed from accurate data by \citet{2019ApJ...870L..10M}. We do not attempt to produce a precise fit to the data, instead, we want to highlight the clear presence of wiggles in the MW's RC. Notice that the vertical amplitude and radial spacing of such wiggles are comparable in magnitude (~10 - 20kms$^{-1}$) to the wiggles in the simulated RC.

\subsection{Discreteness in orbital angular momentum}

Notice that in the $V_{\phi}-R$ plane of an axisymmetric disk, families of constant angular momentum $L_z$ are represented by $V_{\phi} = L_z/R$, i.e., lines that cross diagonally such plane. Those curves of constant $L_z$ would cover homogeneously the $V_{\phi}-R$ plane of an axisymmetric disk. But that is not the case for our model nor the DR2. Because of the bar and spiral arms Figures \ref{ridges1} and \ref{ridges}, and figure 2 in \citet{2018arXiv180410196A}, reveal that the kinematics of the stellar disk is far from being homogeneous. Aside from producing wiggles in the RC, these inhomogeneities reveal more about the dynamics going on in the disk. 

\begin{figure}
\begin{center}
\includegraphics[width=8cm]{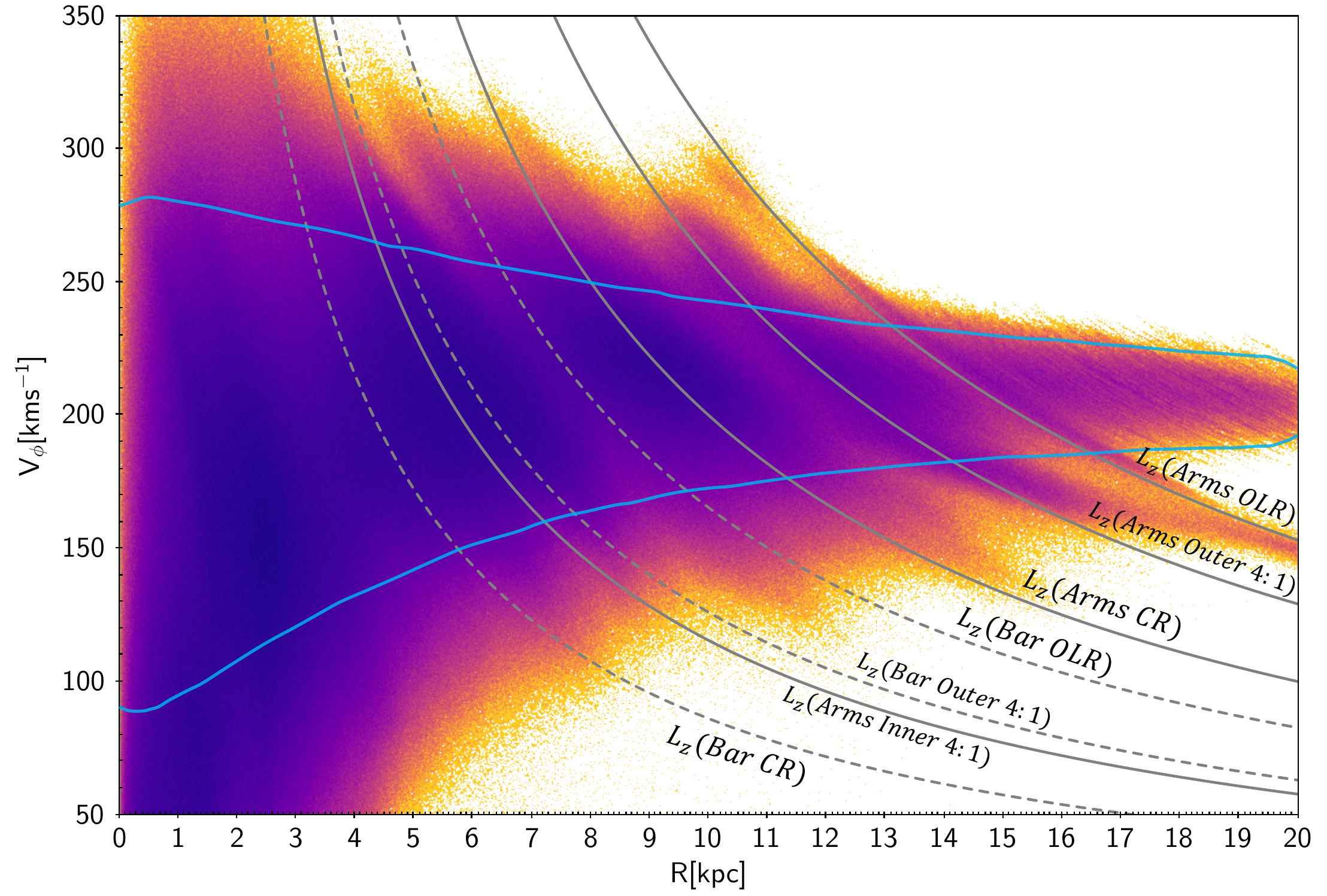}
\end{center}
\caption{Rotation velocity for stars in our MW simulation with bar+spirals. The blue curve is the envelope of the distribution of stars for the axisymmetric case. Grey dashed and solid lines are $L_z$ curves associated with bar and spirals resonances, respectively.}
\label{ridges2}
\end{figure}

In Figure \ref{ridges2} we present again the $V_{\phi}-R$ plane, and for comparison the blue line indicates the envelope of the distribution of stars in the same model but without bar and spirals. As expected, the stellar velocities spread out of the blue envelope, i.e., bar and spirals increase the velocity dispersion $\sigma_{\phi}$ of the disk. But what is interesting is how the spread out of $V_{\phi}$ happens, it is not homogeneous, stars propagate more efficiently along the two extrema of the diagonal ridges. Moreover, we notice that the extrema of each ridge, that stand out as spikes in the $V_{\phi}$ distribution, are connected by a curve of constant angular momentum.

We found that such $L_z$ values are not arbitrary. From the pattern speeds and frequencies of the model, $\Omega(R)$, we know the location $R_{res}$ of bar and spirals resonances, such that the values of $L_z=R_{res}^2\Omega$ can be computed for each resonance. Figure \ref{ridges2} shows the $L_z$ curves associated with each resonance, which layout on the most prominent ridges in the $V_{\phi}-R$ plane. Bar and spirals produce their own ridges, and some of them overlap when the resonances are close enough, as is the case for the Bar's outer 4:1 and Arms inner 4:1 resonances. Also, notice that the prominent ridge that develops in the far right of the distribution is due to the Arms OLR with and important contribution of the one due to the Arms 4:1 resonance, giving a clear signature of resonance overlapping.

Bar + spirals rearrange the stellar kinematics of the stellar disk in such way that stars are placed in orbits with preferential values of angular momentum. The imprint of this process is clear in our model and observed with high resolution within the volume covered by Gaia.

\subsection{Age and metallicity of the ridges}

\begin{figure}
\begin{center}
\includegraphics[width=8.5cm]{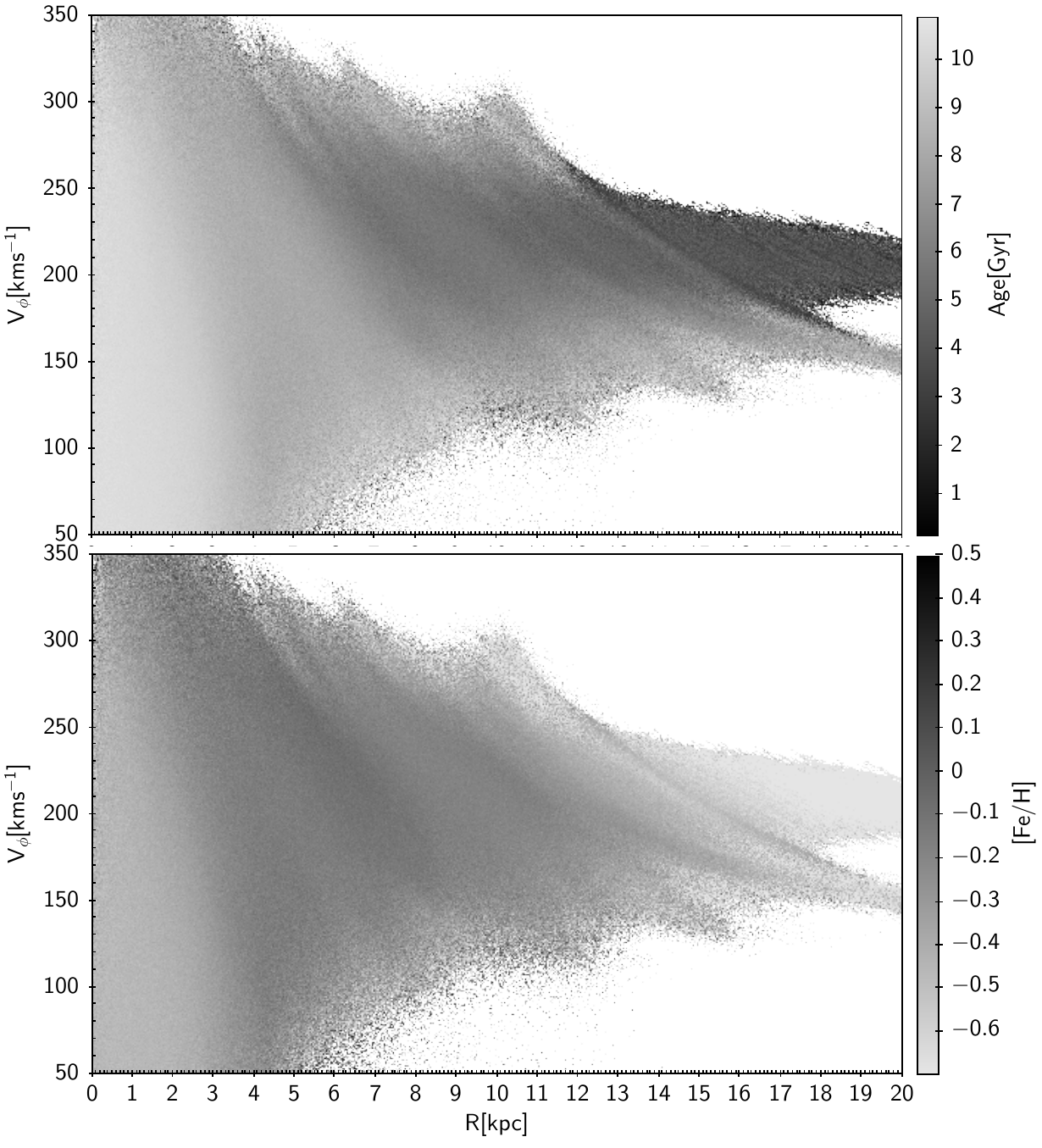}
\end{center}
\caption{Distribution of stars in the $V_{\phi}-R$ plane. The color coding indicates: the mean age at every point in the plane (Top), and the mean metallicity at every point in the plane (Bottom). }
\label{ridges_age}
\end{figure}

The footprint of bar and spiral arms in the stellar velocity distribution (Fig. \ref{ridges}) not only depends on the nature and characteristics of both perturbers but also on the time the stars interact with them, i.e., on the age of the stars. The simulation analyzed here, specifically tailored for the Milky Way, assigns an age label to each star based on the time they are introduced to the simulation. This allows us to split the disk into populations according to the stellar age. Figure \ref{ridges_age} (top panel) shows the $V_{\phi}-R$ plane colored with the average stellar age. A first comparison between Figures \ref{ridges} and \ref{ridges_age} (top) reveals that the diagonal over-densities that give rise to wiggles in the RC are better sampled by young and middle-age stars. On the other hand old stars, with higher velocity dispersion, populate the ends of the ridges.   
We have seen that the diagonal layout of the ridges in the $V_{\phi}-R$ plane naturally gives rise to wiggles in the RC. Figure \ref{ridges_age} shows that for the old stellar population the extrema of the ridges have a larger vertical separation than those for the young population; which means that the wiggles in the RC are more prominent for an old population and have a larger vertical amplitude, being more subtle for younger stars. This is consistent with recent estimations of the MW rotation curve based on GAIA DR2 \citep{Kawata_RC}, where the wiggles are clearly observed although not discussed at all, however the dependence on stellar tracer age is strikingly noticeable.

It is expected for the imprints in the kinematics of the disk to be accompanied by alterations in other observables, for example, radial migration and heating have the potential to shape the local metallicity distribution \citep{Hayden2015,2017MNRAS.468.3615M}. In Figure \ref{ridges_age} (bottom panel) we present again the $V_{\phi}-R$ plane but now colored with the average metallicity. The pattern of ridges found here is dominated by high metallicities, and the whole extension of the ridge is roughly metal-rich.

From Figure \ref{ridges_age} we have learned some more details about the ridges in the stellar velocity distribution. There is a high metallicity component seen all across the ridges in the bottom panel of Figure \ref{ridges_age}; this is an indication that the ridges include a large quantity of stars that originated in the inner disk, whose orbits now reach the outer disk. On the other hand the young component seen in the ridges in the top panel of Figure \ref{ridges_age} does not seem to reach the limits of the ridges; this is an indication that the ridges include young objects with moderate eccentricity as well as older objects with extreme eccentricity.

\section{Summary}
\label{summary}

 The Gaia DR2, with its high resolution and wide spatial coverage, has unveiled a richness of substructures in the kinematics of the MW disk. Of particular interest is the distinctive pattern of diagonal ridges covering the $V_{\phi}-R$ plane \citep{2018arXiv180410196A,2018MNRAS.479L.108K}. Such arrangement of the stellar kinematics is likely to be the signal of  perturbations due to the bar or the spiral arms.  
 
 Here, with a MW Galaxy model that harbours bar + spiral arms, we found not only wide denser ridges, but also many thin ridges, similar to what Gaia has revealed \citep[figure 2 in][]{2018arXiv180410196A}. Given the similarity of Figure \ref{ridges1} with the substructures in the DR2 we can confirm that the duo bar+spirals plays a leading role in perturbing the stellar disk.
 
 Moreover, as anticipated from the kinematics in the DR2, the pattern of ridges is present along all the extension of the disk (Figure \ref{ridges}): from inside the bar region to the outer disk, beyond the extension of the arms. Here we notice that such a large-scale arrangement of the kinematics induced by arms and bar must be visible not only locally, but on global observables of the disk. The layout of the substructures in the $V_{\phi}-R$ plane induce a sequence of maxima and minima in the mean rotation. Hence, we show that, due to their diagonal nature, the ridges manifest themselves as bumps and wiggles in the rotation curve. 
 
 Rotation curves of external galaxies present similar wiggles traced by gas kinematics and their origin has puzzled different studies, with explanations ranging from spiral arms \citep{Paris86} to features in the gravitational potential helping to assess the dark matter nature \citep{Chakra-Skivie2018} or halo importance \citep{DeBlokMacG98}.  Recent Milky Way RC estimations present similar wiggles in the young stellar component \citep[e.g.][]{Kawata_RC}, based on that our study seems to favor the spiral-arm origin even in external galaxies traced by gas, however it is important to clarify that it does not rule out other mechanisms.  
 
 Finally, such diagonal-like layout of overdensities in the $V_{\phi}-R$ plane can be explained as stellar orbits being piled up around curves of constant angular momentum. Which means that wiggles in RCs, as projection of the ridges, are a manifestation of the bar and spirals placing the stars onto orbits with preferential values of angular momentum.

\section*{Acknowledgements}
We would like to acknowledge the anonymous referee for a careful review and insightful suggestions.
We acknowledge DGTIC-UNAM for providing HPC resources on the Supercomputer
Miztli. LAMM is supported by the European Research Council (ERC StG-335936, CLUSTERS). B.P. and LAMM acknowledge support from PAPIIT IN-101918 and IN-105916. B.P. acknowledges CONACYT Ciencia B\'asica grant 255167. A.P. acknowledges support from PAPIIT IN-109716 grant. O.V acknowledges support from PAPITT grant IN-112518.

%\bsp	% typesetting comment
\label{lastpage}

\begin{thebibliography}{101}

\bibitem[Allen \& Santill\'an(1991)]{AS91}
Allen C., Santill\'an A., 1991, \rmxaa, 22, 255

\bibitem[Antoja et al.(2009)]{2009ApJ...700L..78A} Antoja, T., Valenzuela, O., Pichardo, B., et al.\ 2009, \apjl, 700, L78 

\bibitem[Antoja et al.(2014)]{2014A&A...563A..60A} Antoja, T., Helmi, A., Dehnen, W., et al.\ 2014, \aap, 563, A60 

\bibitem[Antoja et al.(2018)]{2018arXiv180410196A} Antoja, T., Helmi, A., Romero-G{\'o}mez, M., et al.\ 2018, \nat, 561, 360 

\bibitem[Carigi \& Peimbert(2011)]{2011RMxAA..47..139C} Carigi, L., \& Peimbert, M.\ 2011, \rmxaa, 47, 139.

\bibitem[Chakrabarty \& Sikivie(2018)]{Chakra-Skivie2018} Chakrabarty, S.~S., \& Sikivie, P.\ 2018, \prd, 98, 103009 

\bibitem[de Blok \& McGaugh(1998)]{DeBlokMacG98} de Blok, W.~J.~G., \& McGaugh, S.~S.\ 1998, \apj, 508, 132 

\bibitem[Dehnen(2000)]{2000AJ....119..800D} Dehnen, W.\ 2000, \aj, 119, 800

\bibitem[Fragkoudi et al.(2019)]{2019arXiv190107568F} Fragkoudi, F., Katz, D., White, S.~D.~M., et al.\ 2019, arXiv e-prints , arXiv:1901.07568.

\bibitem[Freudenreich(1998)]{1998ApJ...492..495F} Freudenreich, H.~T.\ 1998, \apj, 492, 495.

\bibitem[Hattori et al.(2019)]{2018arXiv180401920H} Hattori, K., Gouda, N., Tagawa, H., et al.\ 2019, \mnras, 484, 4540.

\bibitem[Hayden et al.(2015)]{Hayden2015} 
Hayden, M.~R., Bovy, J., Holtzman, J.~A., et al.\ 2015, \apj, 808, 132

\bibitem[Hunt et al.(2018)]{2018MNRAS.481.3794H} Hunt, J.~A.~S., Hong, J., Bovy, J., Kawata, D., \& Grand, R.~J.~J.\ 2018, \mnras, 481, 3794 

\bibitem[Kalnajs(1983)]{Kalnajs1983}
Kalnajs, A.J., 1983, in Internal Kinematics and Dynamics of Galaxies, IAU Symposium Vol. 100, edited by E. Athanassoula (Reidel, Dordrecht), p. 87.


\bibitem[Kawata et al.(2018)]{2018MNRAS.479L.108K} Kawata, D., Baba, J., Ciuc{\v a}, I., et al.\ 2018, \mnras, 479, L108 

\bibitem[Kawata et al.(2019)]{Kawata_RC} Kawata, D., Bovy, J., Matsunaga, N., \& Baba, J.\ 2019, \mnras, 482, 40

\bibitem[Kent(1986)]{1986AJ.....91.1301K} Kent, S.~M.\ 1986, \aj, 91, 1301

\bibitem[Khoperskov et al.(2018)]{2018arXiv181109205K} Khoperskov, S., Di Matteo, P., Gerhard, O., et al.\ 2018, arXiv:1811.09205 

\bibitem[Laporte et al.(2019)]{2018arXiv180800451L} Laporte, C.~F.~P., Minchev, I., Johnston, K.~V., et al.\ 2019, \mnras, 485, 3134. 

\bibitem[Li et al.(2019)]{2019MNRAS.482.5106L} Li, P., Lelli, F., McGaugh, S.~S., Starkman, N., \& Schombert, J.~M.\ 2019, \mnras, 482, 5106 


\bibitem[Martinez-Medina et al.(2017)]{2017MNRAS.468.3615M} Martinez-Medina, L.~A., Pichardo, B., Peimbert, A., \& Carigi, L.\ 2017, \mnras, 468, 3615 


\bibitem[Mr{\'o}z et al.(2019)]{2019ApJ...870L..10M} Mr{\'o}z, P., Udalski, A., Skowron, D.~M., et al.\ 2019, \apj, 870, L10..

\bibitem[Palunas, \& Williams(2000)]{2000AJ....120.2884P} Palunas, P., \& Williams, T.~B.\ 2000, \aj, 120, 2884.

\bibitem[P{\'e}rez-Villegas et al.(2017)]{2017ApJ...840L...2P} P{\'e}rez-Villegas, A., Portail, M., Wegg, C., \& Gerhard, O.\ 2017, \apjl, 840, L2

\bibitem[Persic et al.(1996)]{1996MNRAS.281...27P} Persic, M., Salucci, P., \& Stel, F.\ 1996, \mnras, 281, 27.

\bibitem[\protect\citeauthoryear{Pichardo et al.}{2003}]{PMME03}
  Pichardo, B., Martos, M., Moreno, E. \& Espresate, J., 2003, ApJ,
  582, 230

\bibitem[Pichardo et al.(2004)]{Pichardo2004} Pichardo, B., Martos, 
M., \& Moreno, E.\ 2004, \apj, 609, 144

\bibitem[Pismis(1986)]{Paris86} Pismis, P.\ 1986, \rmxaa, 12, 79 

\bibitem[Quillen et al.(2018)]{2018MNRAS.480.3132Q} Quillen, A.~C., Carrillo, I., Anders, F., et al.\ 2018, \mnras, 480, 3132 

\bibitem[Ramos et al.(2018)]{2018arXiv180509790R} Ramos, P., Antoja, T., \& Figueras, F.\ 2018, \aap, 619, A72.

\bibitem[Reid et al.(2014)]{2014ApJ...783..130R} Reid, M.~J., Menten, K.~M., Brunthaler, A., et al.\ 2014, \apj, 783, 130.

\bibitem[Sellwood et al.(2019)]{2019MNRAS.484.3154S} Sellwood, J.~A., Trick, W.~H., Carlberg, R.~G., et al.\ 2019, \mnras, 484, 3154. 

\end{thebibliography}
\end{document}